# AI-Based Impedance Encoding-Decoding Method for Online Impedance Network Construction of Wind Farms

Xiaojuan Zhang
*School of Electrical Engineering*
*Shanghai Jiao Tong University*
Shanghai, China
xiaojuanzhang123@gmail.com

Tianyu Jiang
*School of Electrical Engineering*
*Shanghai Jiao Tong University*
Shanghai, China
jty030522@sjtu.edu.cn

Haoxiang Zong
*School of Electrical Engineering*
*Shanghai Jiao Tong University*
Shanghai, China
haoxiangzong@sjtu.edu.cn

Chen Zhang
*School of Electrical Engineering*
*Shanghai Jiao Tong University*
Shanghai, China
nealbc@sjtu.edu.cn

Chendan Li
*Department of Naval, Electrical, Electronic and Telecommunications Engineering*
*University of Genova*
Genova, Italy
chendan.li@diten.unige.it

Marta Molinas
*Department of Engineering Cybernetics*
*Norwegian University of Science and Technology*
Trondheim, Norwegian
marta.molinas@ntnu.no

*Abstract*—The impedance network (IN) model is gaining popularity in the oscillation analysis of wind farms. However, the construction of such an IN model requires impedance curves of each wind turbine under their respective operating conditions, making its online application difficult due to the transmission of numerous high-density impedance curves. To address this issue, this paper proposes an AI-based impedance encoding-decoding method to facilitate the online construction of IN model. First, an impedance encoder is trained to compress impedance curves by setting the number of neurons much smaller than that of frequency points. Then, the compressed data of each turbine are uploaded to the wind farm and an impedance decoder is trained to reconstruct original impedance curves. At last, based on the nodal admittance matrix (NAM) method, the IN model of the wind farm can be obtained. The proposed method is validated via model training and real-time simulations, demonstrating that the encoded impedance vectors enable fast transmission and accurate reconstruction of the original impedance curves.

## I. Introduction

With the increasing penetration of the converter-interfaced renewable energy base, wideband oscillations induced by converter's control dynamics are prone to occur, which greatly hinder the stable grid integration [1]. To effectively prevent the oscillation, the system stability margin and the potential oscillation source need to be acquired for the proper activation of the damping control. Such information is affected by all generation units within the farm and their time-varying operation points, and thus an online oscillation assessment that can incorporate multi-machine dynamics and their real-time operating states is required.

In this context, the IN model-based oscillation diagnosis method can serve as a good tool [2], due to the advantages of compactness, measurability, circuit scalability, etc. The basic idea of such method is to first establish the impedance model of each generation unit containing both control dynamics and operation points, e.g., *dq* impedance [3], modified sequence impedance [4], etc.; Then, the impedance model of each unit will be assembled with the impedance of transmission lines to obtain the IN model of the whole farm, based on the nodal admittance or loop-gain modeling method [5], [6]. Since the established IN model retains the structural information of the whole farm (e.g., nodes, branches), time-domain or frequency-domain modal analysis methods can be applied to determine the location of the potential oscillation source by calculating the modal sensitivity information, e.g., node participation factors (PF) and parameter sensitivity [7]-[9].

The above analysis procedure is effective offline, however, for its online application, the calculation speed of such high-dimensional IN model needs to be considered. Recently, artificial intelligence (AI) methods have been adopted to accelerate the impedance calculation process by establishing a mapping relationship from the generation unit's operating point to the impedance curve [10]-[12]. However, these studies are mainly aimed for the single machine system, and how to acquire the IN model based on all units' impedance curves are rarely discussed. Since each impedance curve will span a wide frequency range, the direct transmission of all units' high-density impedance curves to the farm-side will significantly slow down its oscillation assessment efficiency.

To tackle with the above issue of the online application, this paper proposes an AI-based impedance encoding-decoding method, which will compress impedance data for the fast transmission and then reconstruct them at the farm-side for the accurate assessment. The method is tested on a dataset of 100 impedance curve samples that is used for the encoder-decoder model training, and a t-SNE visualization of the latent impedance vectors is conducted to verify the effectiveness of the dimensionality reduction. At last, the online performance of the method is validated using a typical wind farm on a real-time simulation platform.

## II. Impedance Encoding and Decoding Method for Online Impedance Network Construction

### A. Overall Architecture of the Method

The overall architecture of the proposed impedance encoding and decoding method is presented in Fig. 1. Firstly, the impedance curve of each wind turbine is generated locally and refreshed according to their time-varying operating states. Secondly, the high-density impedance curve (e.g., 1-2500Hz) will be passed to the impedance coder at turbine-side, where it is compressed into a much more compact representation.

This paper was supported by National Natural Science Foundation of China (No. 52207215) and the Delta Power Electronics Science and Education Development Program of Delta Group (No. DREK2023004).

Such compressed data will be transmitted to the farm-side via the embedded communication network, such as the SCADA system. Finally, the farm-side gathers the transmitted data from all turbines and utilizes the pre-trained impedance decoder to reconstruct the original impedance curves. Based on the nodal admittance matrix modeling method, the closed-loop IN model can be obtained for the oscillation diagnosis.

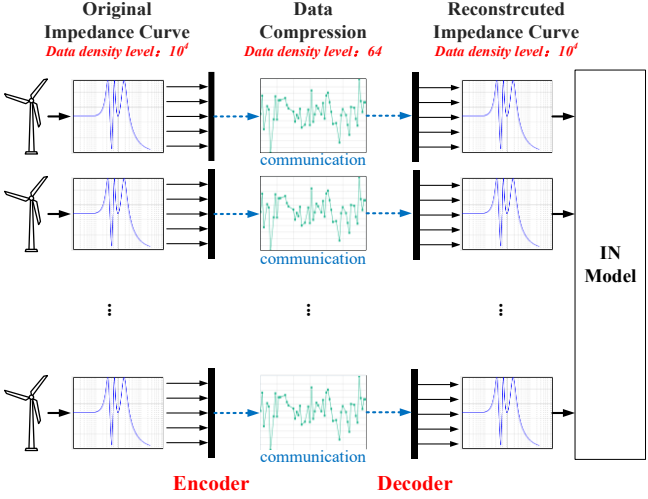

Fig. 1. The architecture of the impedance encoding-decoding method for the online impedance network construction

### B. MLP-based Impedance Encoding and Decoding Method

To address the challenge of efficiently transmitting wideband, high-resolution impedance curves from multiple wind turbines to the farm-side, a fully connected autoencoder architecture is proposed based on two multilayer perceptron (MLPs). The model compresses each impedance sample into a compact latent representation for communication efficiency and reconstructs the original impedance profile with minimal distortion. The designed autoencoder comprises two primary components: (a) an *encoder* that performs dimensionality reduction; (b) a *decoder* that reconstructs the full impedance data from the latent representation. The entire model is trained in an end-to-end fashion using a relative reconstruction loss. The overall architecture is illustrated in Fig. 2.

**Encoder Module**: Let each impedance sample be represented by a two-by-two matrix, e.g., *dq* impedance [3].

$$Y_i = \left\{Y_i^{(t)} \in C^{2\times 2}\right\}_{t=1}^{T}, \quad (1)$$

where $T$ is the number of frequency points. Since the matrix elements of the *dq* admittance is in complex form, the *dq* admittance is vectorized into its four matrix elements and then decomposed into real and imaginary parts, resulting in a real-valued input tensor, which is then subsequently flattened (assuming $T$=2500.).

$$x_i \in R^{2\times 4\times T} \Rightarrow x_i \in R^{20000} \quad (2)$$

The encoder maps the high-dimensional input $x_i$ to a low-dimensional latent vector $h_i \in R^d$ through a series of nonlinear transformations:

$$h_i = \phi_3(W_3 \cdot \phi_2(W_2 \cdot \phi_1(W_1 \cdot x_i + b_1) + b_2) + b_3) \quad (3)$$

where $\phi_k(\cdot)$ denote activation function; $W_1$, $W_2$, $W_3$ represent the learnable weight parameters; $b_1$, $b_2$, $b_3$ represent the learnable bias parameters. The encoder layer dimensions are: 20000→2048→512→64.

**Decoder Module**: The decoder reverses the compression process and reconstructs the impedance curve:

$$\hat{x}_i = \psi_3(V_3 \cdot \psi_2(V_2 \cdot \psi_1(V_1 \cdot h_i + c_1) + c_2) + c_3), \quad (4)$$

where $\psi_k(\cdot)$ represents the activation function, and $V_1$, $V_2$, $V_3$ represent the learnable weight parameters of the decoder. The output $\hat{x}_i \in R^{20000}$ is reshaped back to the original format (T, 2, 4) to yield the estimated impedance curve.

Both encoder and decoder modules use the Rectified Linear Unit (ReLU) activation function, defined as:

$$\phi(h) = max(0, h), \quad (5)$$

which introduces non-linearity and ensures positive activations, aiding in the learning of complex mappings between input and latent space.

**Loss Function:** To enhance the robustness and ensure the generalization capability across impedance curves of varying magnitudes, a relative error-based reconstruction loss is adopted here. For each sample, the loss is computed as:

$$\mathcal{L}_{recon(i)} = \frac{|\hat{x}_i - x_i|_2}{|x_i|_2 + \epsilon}, \quad (6)$$

where $\epsilon = 10^{-8}$ is added for numerical stability.

The total training loss is defined as the average relative reconstruction error over all N training samples:

$$\mathcal{L}_{total} = \frac{1}{N}\sum_{i=1}^{N}\mathcal{L}_{recon(i)}. \quad (7)$$

It is worth of noting that the entire training process is conducted in an unsupervised manner, requiring no labeled data. This greatly reduces the cost and complexity of data preparation, making the proposed method particularly suitable for transmitting huge amount of impedance data from multiple wind turbines in a large-scale wind farm. These advantages underscore the practicality and scalability of the proposed framework.

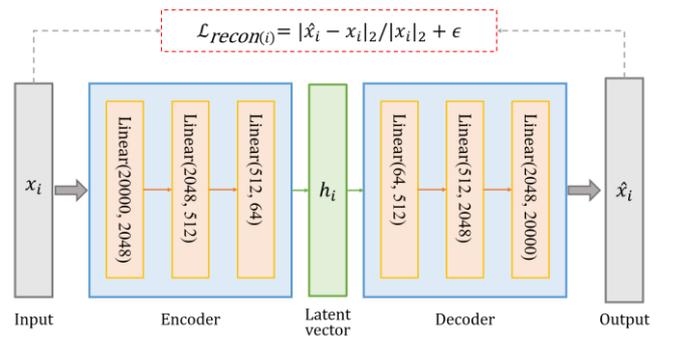

Fig. 2. Structure of the MLP-based Encoding and Decoding. Each input data $x_i \in R^{20000}$, obtained by flattening the frequency-domain impedance matrix over all frequency points, is first passed through an encoder composed of three fully connected layers. The encoder outputs a low-dimensional latent vector $h_i$. The decoder, symmetric to the encoder, reconstructs the impedance data $\hat{x}_i$ from $h_i$.

### C. Establishment of the IN Mode

According to [5], the closed-loop stability analysis model of a typical wind farm can be represented by its *s*-domain nodal admittance matrix $Y_{node}(s)$ as:

$$Y_{\text{node}}(s) = Y_{\text{wt}}(s) + Y_{\text{net}}(s) \tag{8}$$

where the $Y_{\text{wt}}(s)$ denotes the admittance matrix with diagonal elements as each turbine's admittance:

$$Y_{\text{wt}}(s) = \begin{bmatrix} Y_{\text{wt1}}(s) & & \\ & \ddots & \\ & & Y_{\text{wtn}}(s) \end{bmatrix} \tag{9}$$

and $Y_{\text{net}}(s)$ is the admittance matrix of the collection network, which can be obtained using the incidence matrix method [5]. Based on such model retaining the structural information of the wind farm, the stability margin and the oscillation source analysis can be carried out.

## III. Case Study

### A. Test System

In this work, a simulation model of a grid-tied wind farm system with four turbines as shown in Fig. 3 was established as the test system. As to the simulation, each wind turbine is represented by a simplified VSC with phase-locked loop (PLL) and current loop control, where the rated dc voltage is 1500V and ac voltage is 690V. All the four turbines are connected to a 35kV bus through transformers. The operating points of each wind turbine are set differently so that their *dq* admittances vary accordingly. Meanwhile, the ac grid is modeled using a Thevenin equivalent circuit. The simulation model runs on a real-time simulator MT 8020-16, and the compressed impedance data obtained in the simulator was sent to the host computer, which will be reconstructed for the establishment of the complete IN model.

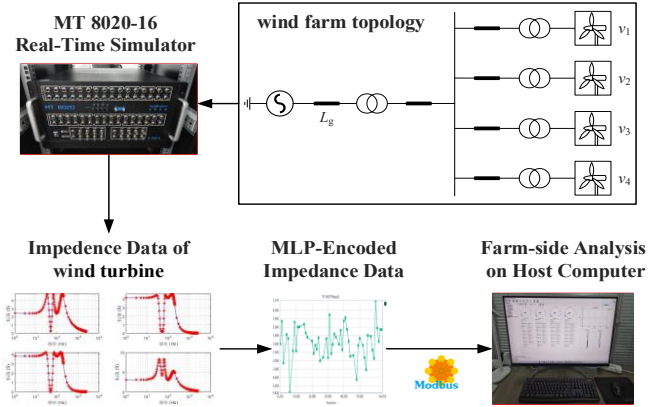

Fig. 3. The test wind farm and real-time simulation platform

### B. Validation of the Impedance Encoding and Decoding

#### 1) Validation of the Training Results

**Dataset and Training Setup:** The whole dataset consists of 100 samples, each representing a 2×2 complex impedance matrix evaluated at 2500 frequency points. For each sample, the real and imaginary components are extracted and flattened into a 1×20000 vector to serve as the input to the network. The data is globally normalized using z-score as:

$$x_{\text{norm}} = \frac{x - \mu}{\sigma}, \tag{10}$$

where $\mu$ and $\sigma$ are the mean and standard deviation across the entire dataset. 90 samples are used for training and 10 for the testing. The model is trained using the Adam optimizer with a learning rate of $1\times10^{-4}$, a batch size of 10, and for 500 epochs. Training is conducted on a GPU-enabled machine with automatic saving of the model state at each epoch.

**Results and Generalization:** Fig. 4 shows the training loss curve over 500 epochs. The relative error steadily decreases and eventually converges to a stable value around 0.0204, indicating that the model successfully captures the underlying patterns in the impedance data.

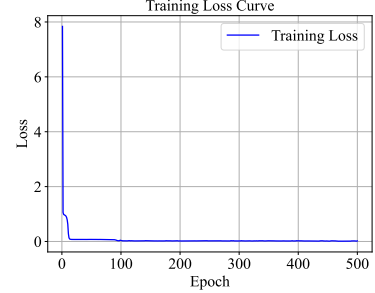

Fig. 4. The training loss curve.

To evaluate the generalization capability of the trained autoencoder, we perform both in-sample (training set) and out-of-sample (testing set) reconstructions. For each sample, the original and reconstructed impedance matrices are visualized in both:

- Amplitude domain: $A(f)=|Y(f)|$
- Phase domain: $\phi(f)=\angle Y(f) \cdot 180/\pi$

The results (see Fig. 5) show that: The reconstructed amplitudes closely follow the original curves, including resonant peaks and low-frequency characteristics. Phase curves show slightly larger discrepancies, especially in high-frequency ranges, due to their lower signal-to-noise ratio. No visible overfitting is observed, as reconstruction performance on the test set is comparable. This validates the generalization ability and robustness of the trained model across a wide range of operating conditions.

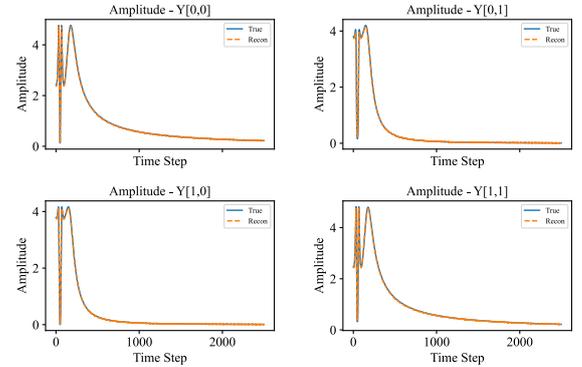

(a) Results of amplitude for testing

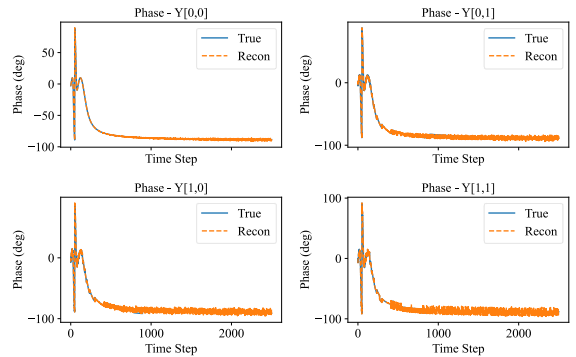

(b) Results of phase for testing

Fig. 5. Comparison of true and reconstructed curves.

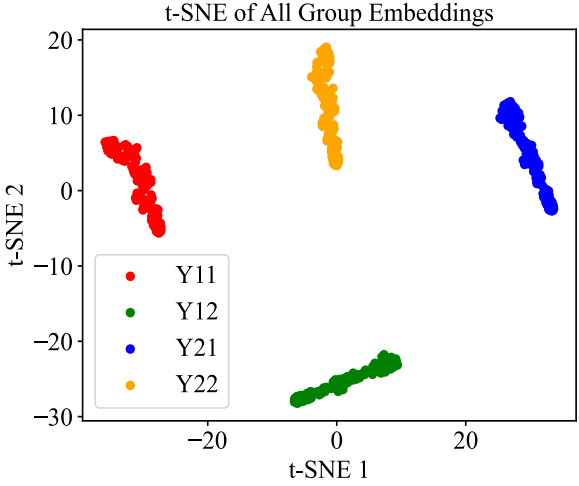

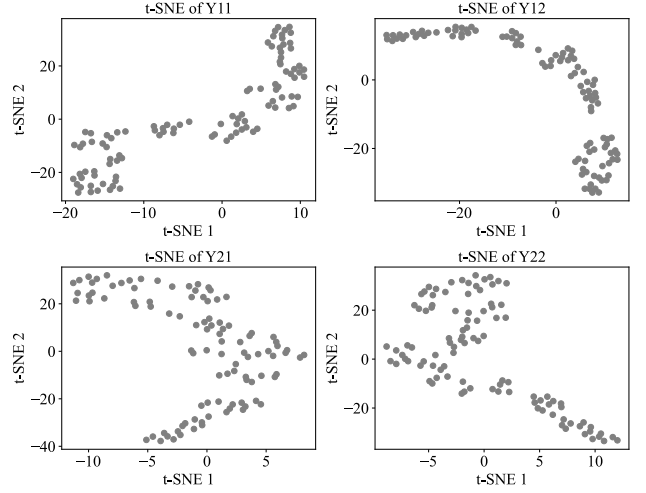

Fig. 6. t-SNE visualization of all semantic group embeddings from the latent space.

Fig. 7. t-SNE visualization of latent representations corresponding to each impedance matrix element.

*2) Latent Vector Analysis*

To examine whether the proposed MLP encoder preserves element-wise semantic information in the latent representation, a set of visualization experiments on the 64-dimensional latent vectors is performed. Each vector is produced from the impedance curve of a single wind turbine, and its internal structure is explicitly designed to correspond to the elements of the original 2×2 impedance matrix.

**Experimental Setup:** For each sample, the impedance matrix $Y(f) \in C^{2\times 2}$ is extracted across 2500 frequency points. Each element (i.e., $Y_{11}, Y_{12}, Y_{21}, Y_{22}$) is a frequency-dependent complex value. During the preprocessing stage, the 20000-dimensional input vector is reduced to a 64-dimensional latent vector. The latent vector is divided into four equal-sized semantic groups:
- Dimensions 1–16 encode features of $Y_{11}$
- Dimensions 17–32 encode features of $Y_{12}$
- Dimensions 33–48 encode features of $Y_{21}$
- Dimensions 49–64 encode features of $Y_{22}$

Each group is intended to preserve information related to its corresponding impedance matrix element, and all four are considered equally important. We apply the t-distributed Stochastic Neighbor Embedding (t-SNE) algorithm [13] to visualize the structure of these semantic groups in a 2D plane.

**Results on Joint Embedding of All Groups:** As shown in Fig. 6, each point represents one semantic group (i.e., one element of the impedance matrix) from one turbine sample. All 400 vectors (100 samples×4 elements) are projected into 2D using t-SNE, and colored according to their associated matrix element. The result reveals a clear separation among the four groups. The embeddings of $Y_{11}, Y_{12}, Y_{21}, Y_{22}$ form distinct and non-overlapping clusters in the latent space. This confirms that the encoder successfully learns to encode each impedance element independently, with minimal semantic entanglement or redundancy.

**Results on Independent Embedding of Each Group:** In Fig. 7, the latent representations of each group individually across all samples are further visualized. For each matrix element, its 16-dimensional latent sub vectors are projected to 2D using t-SNE. The results show smooth and well-organized manifolds for all four groups. No significant outliers or abnormal scattering patterns are observed. This suggests that for each impedance matrix element, the encoder produces consistent and structured latent embeddings across different turbine operating conditions.

*C. Online Application of the Proposed Method*

The four-wind-turbine system runs on a real-time simulator and the compressed latent vectors are send to the host computer and visualized, as shown in Fig. 8. the time-domain waveform of active power $P$, power factor $\cos\varphi$, and nominal value of the voltage at the grid point $U_{\text{pcc}}$ are plotted in Fig. 8(a). Meanwhile, the curves in Fig. 8(b) correspond to the compressed latent vector of each turbine. It can be observed that the latent vectors exhibit a consistent overall trend, while retaining distinguishable variations in certain dimensions. This indicates that the proposed model is capable of preserving shared structural characteristics while capturing turbine-specific features in the latent space.

Furthermore, the above compressed latent vectors are used for the impedance reconstruction via the pre-trained decoder. As in Fig. 9, the reconstructed impedance curves obtained from the decoder closely match the original curves, demonstrating the effectiveness of the proposed method in compressing and reconstructing high-density impedance data.

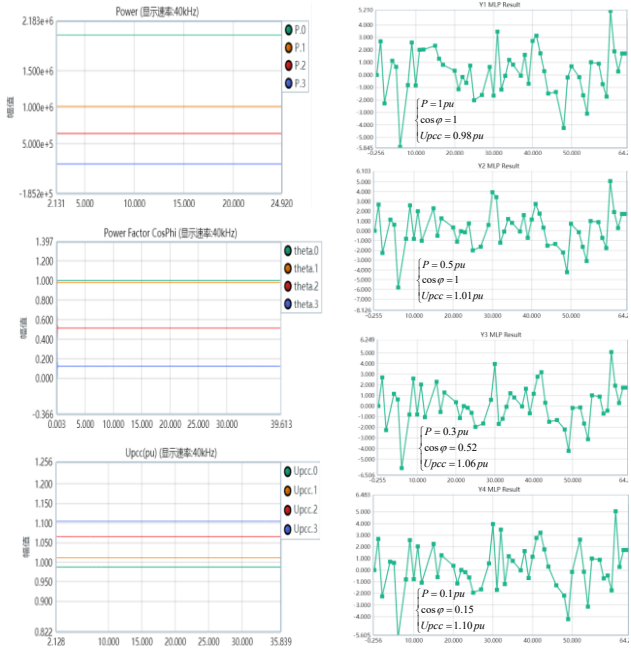
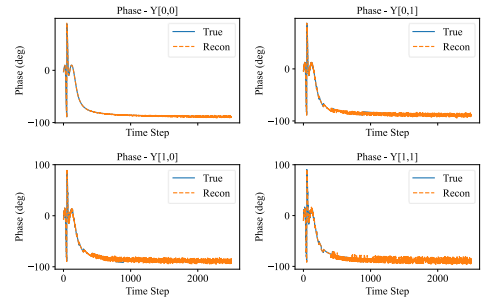

(a) Time-domain waveform of each turbine's operation point

(b) compressed latent vector of turbines

Fig. 8. Visualization of 64-dimensional encoded latent vectors for each wind turbine

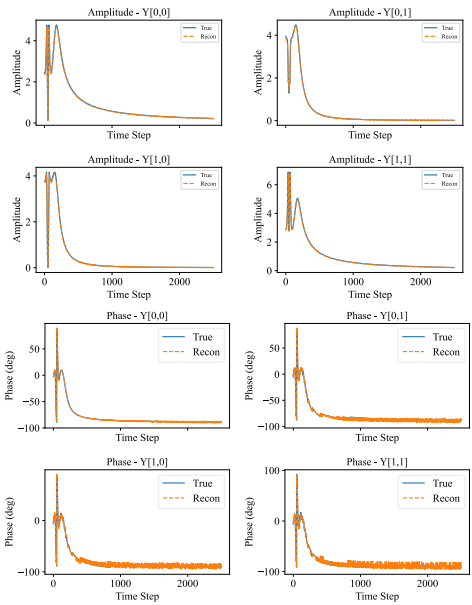

(a) Impedance curves of amplitude and phase for Turbine 1

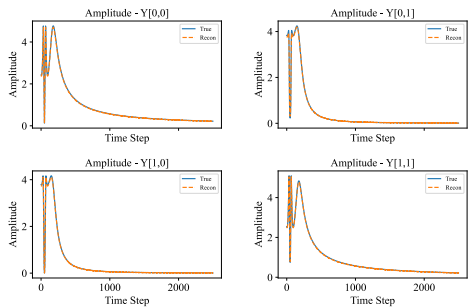

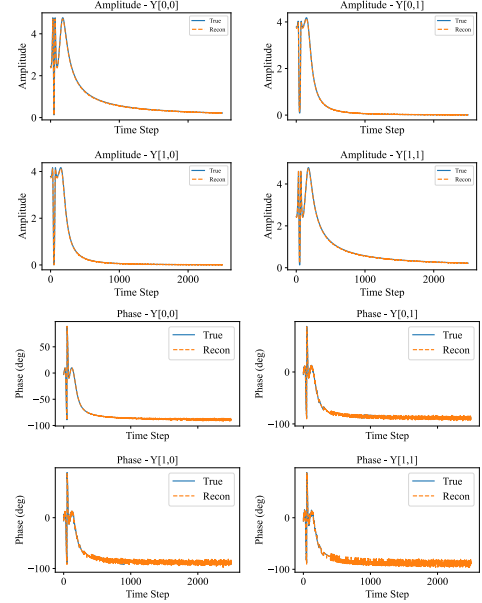

(b) Impedance curves of amplitude and phase for Turbine 2

(c) Impedance curves of amplitude and phase for Turbine 3

(d) Impedance curves of amplitude and phase for Turbine 4

Fig. 9. Simulation results: reconstruction of impedance curves for each wind turbine based on the uploaded compressed data, and comparison with the original impedance curve

## IV. Conclusion

In this paper, an impedance encoding and decoding method based on the multilayer perceptron is proposed, which can greatly reduce the communication burden of uploading the high-density impedance curves, and thus facilitate the online construction of the wind farm impedance network model. The proposed method is validated through convergence and reconstruction analysis, demonstrating the accurate recovery of impedance curves in both amplitude and phase domains. The t-SNE visualization further verifies that the latent space captures structured and disentangled representations for each impedance matrix element. In the future, the authors will further investigate the application of the proposed method for the online evaluation of the system stability margin and oscillation source identification, providing the guidance for the oscillation suppression.